\documentclass{article}
\usepackage{xcolor}

\usepackage{microtype}
\usepackage{graphicx}
\usepackage{subfigure}
\usepackage{booktabs} 

\usepackage{hyperref}

\usepackage[accepted]{icml2021_oppo}
\usepackage[]{icml2021_oppo}

\icmltitlerunning{Over-parameterization and generalization in Audio classification}

\begin{document}

\twocolumn[
\icmltitle{Over-Parameterization and Generalization in Audio Classification}




\begin{icmlauthorlist}
\icmlauthor{Khaled Koutini}{cp}
\icmlauthor{Hamid Eghbal-zadeh}{cp,lit}
\icmlauthor{Florian Henkel}{cp}
\icmlauthor{Jan Schlüter}{cp}
\icmlauthor{Gerhard Widmer}{cp,lit}

\end{icmlauthorlist}

\icmlaffiliation{cp}{Institute of Computational Perception, Johannes Kepler University Linz, Austria}
\icmlaffiliation{lit}{LIT AI Lab}

\icmlcorrespondingauthor{Khaled Koutini}{Khaled.koutini@jku.at}

\icmlkeywords{Machine Learning, ICML}

\vskip 0.3in
]



\printAffiliationsAndNotice{}  

\begin{abstract}
Convolutional Neural Networks (CNNs) have been dominating classification tasks in various domains, such as machine vision, machine listening, and natural language processing. 
In machine listening, while generally exhibiting very good generalization capabilities, CNNs are sensitive to the specific audio recording device used, which has been recognized as a substantial problem in the acoustic scene classification (DCASE) community.
In this study, we investigate the relationship between over-parameterization of acoustic scene classification models, and their resulting generalization abilities. 
Specifically, we test scaling CNNs in width and depth, under different conditions. Our results indicate that increasing width improves generalization to unseen devices, even without an increase in the number of parameters.
\end{abstract}

\section{Introduction}
\label{sec:intro}
An important part of machine auditory perception is to assign tags to audio clips, based on the acoustic activities and events present in the clip. 
This task has various applications such as content-based multimedia information retrieval, context-aware smart devices, and monitoring and surveillance systems.
Convolutional Neural Networks (CNNs) have dominated this area for tasks such as  Acoustic Scene Classification (ASC) and tagging~\cite{hersheyCNNArchitecturesLargescale2017}. 
Building on successful architectures from the vision domain, the CNN architectures used in the audio domain are usually trained using image-like features such as spectrograms that are extracted from the given audio waveforms.
Hence, many deep architectures from vision, such as ResNet~\cite{heDeepResidualLearning2016}, have been adapted to become state-of-the-art models in audio tasks, e.g., 
by regularizing their Receptive Field (RF)~\cite{Koutini2019Receptive}, and Effective Receptive Field (ERF)~\cite{luoUnderstandingEffectiveReceptive2016, koutini21journal}.

\begin{figure}[t]
\begin{center}
\centerline{\includegraphics[width=0.95\columnwidth]{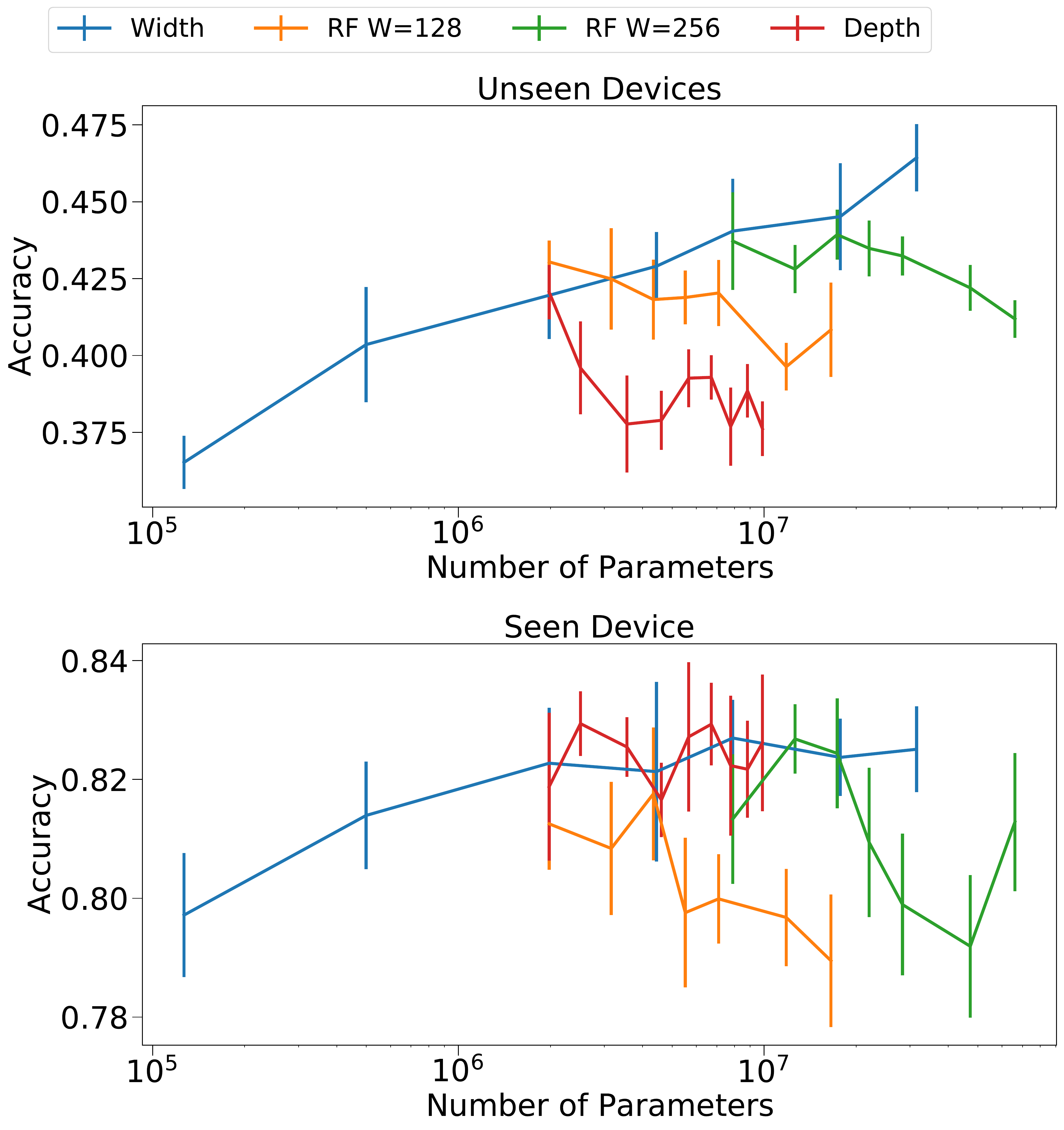}}
\caption{Test accuracy of a Damped ResNet with scaled up variants in width and depth with/without increasing the Receptive Field (RF), with respect to the number of parameters, as explained in Section~\ref{sec:network}.  }
\label{fig:summery:damped}
\end{center}
\vskip -0.3in
\end{figure}
Several studies have shown that scaling up the number of parameters of neural networks improves their performance in different domains, such as vision ~\cite{NeyshaburLBLS19roleofoverparam,TanL19effcientnet} and language~\cite{ Kaplan2020scalinglawslanguage}.
\citet{TanL19effcientnet} investigate the effect of increasing width and depth of CNNs in image classification, and further introduce a high-performance parameter-efficient architecture that exploits the balance between width and depth.
\citet{belkin2019} and \citet{NakkiranKBYBS20ICLR} study the double descent phenomenon and show that when increasing some models' complexities pass the point they can perfectly fit the training data, they enter the over-parameterized regime, where test set performance improves and surpasses the best under-parameterized setting;
even in the presence of limited label noise in training data.
\citet{golubeva2020wider} explore the relation between width and generalization. They discover that by increasing width, generalization can be improved even when the number of parameters is kept fixed by introducing sparsity. 
Although the aforementioned works have shown interesting connections between over-parameterization and generalization, this connection has not yet been studied in the audio domain, and the challenges of designing effective over-parameterized models for acoustic tasks are still unknown.
This becomes an important aspect since the ultimate goal in many applications is to build models that can generalize to new devices, while keeping the complexity low in order to make them suitable for mobile and edge devices.

In this work, we study the characteristics of over-parameterized CNNs in audio classification. We focus on an important generalization aspect: generalization to unseen devices. This is an important and known problem in the acoustic domain, and several international challenges in this area have been dedicated to it.\footnote{\url{http://dcase.community/challenge2020/task-acoustic-scene-classification}}
More specifically, we wish to answer the double question of what scaling approaches (if any) achieve better generalization to unseen devices, and whether the added parameters are the cause of generalization, or a side-effect of scaling. 
To that end, we perform a set of experiments scaling up two architectures that were previously shown to perform well on the ASC task, using four strategies: by 1) width; 2) depth; 3) depth without increasing the RF; and 4) width without increasing the number of parameters. 
The results show that the width of the network plays the biggest role in its generalization to unseen devices, even when the number of parameters are kept fixed.
In contrast, increasing the depth can even hurt generalization. Finally, we will show how these results are connected to the ERF of the scaled up architectures. 


\section{Scaling up CNNs for Audio Classification}

\citet{TanL19effcientnet} observed that, for image classification, scaling up the width and depth of the network, or the input resolution, improves a model's accuracy. 
However, model depth and its relation to generalization, is a different story in deep models using spectrograms for audio tasks.
\citet{koutini21journal} showed that scaling up the depth of a network without restricting its Receptive Field (RF), negatively impacts the generalization capabilities of different CNN architectures.
Further, they proposed several techniques to regularize the RF of models in order to increase the depth, and improve generalization.
\citet{golubeva2020wider} argue that in over-parameterized regimes, 
widening the network while maintaining the number of parameters results in models with better generalization. 

In Section~\ref{sec:network} we first investigate the effect of scaling up the depth; both with, and without changing the maximum RF of the base network.
Furthermore, we compare the performance of scaled-up models; when their width and depth have been increased.

%
%

\section{Aspects of Generalization in Audio}
\label{sec:task}

\begin{figure}[t]
\begin{center}
\centerline{\includegraphics[width=\columnwidth]{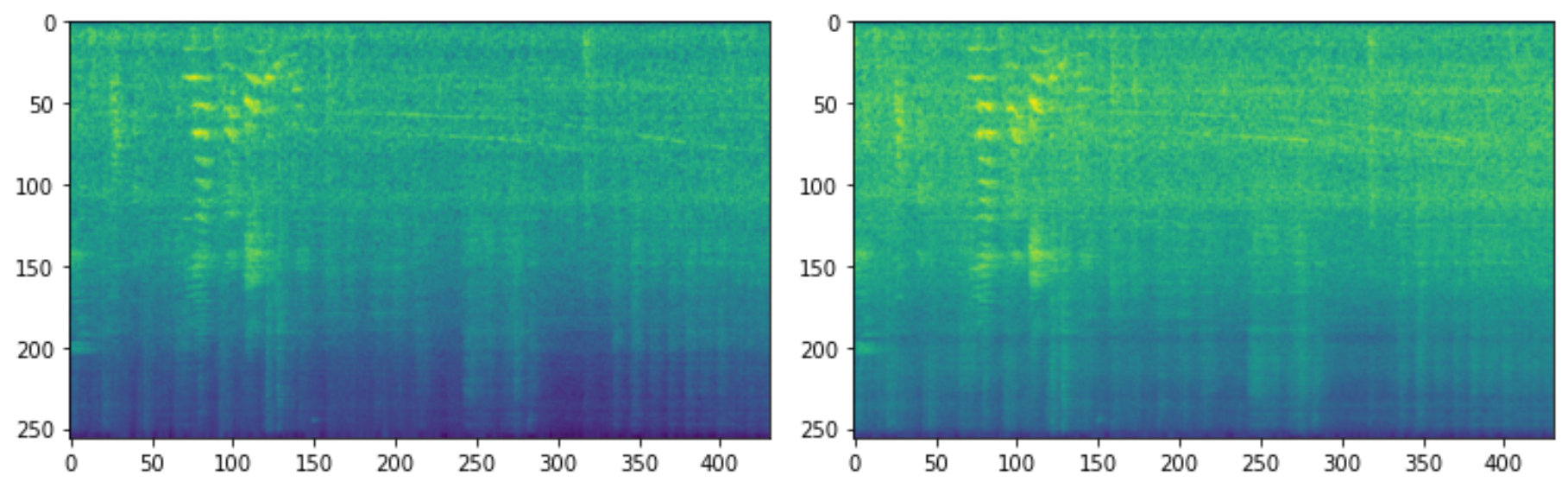}}
\caption{Spectrograms of the same audio recorded with two different devices.}
\label{fig:specs}
\end{center}
\vskip -0.4in
\end{figure}

One of the main problems in audio classification and tagging, is the inability of models to generalize to new devices. 
Using different recording devices at test time introduces a non-trivial domain shift that causes significant performance degradation in tasks such as ASC and tagging. 
Different recording devices may vary in terms of their level of amplitude, the dynamic range, as well as their response to certain frequencies (cf. Figure~\ref{fig:specs}).
The recording device introduces a nonlinear transformation on the input signal, and thereby to the spectrograms calculated from it.
We believe that tasks such as ASC under device mismatch can be a practical test bed for evaluating the generalization capabilities of CNNs.
Hence, in our empirical study, we focus on this task.
%
We use the TAU Urban Acoustic Scenes 2020 Mobile~\cite{Heittola2020} dataset for our analysis, since it is a well-established benchmark for ASC, and additionally, it provides device annotations. The dataset comprises of 64 hours of audio recorded in 10 different environmental scenes -- which are the classification targets -- such as airport, shopping mall, bus etc.  The audio is recorded in 12 European cities using 3 different microphones: Device A ("Soundman OKM II Klassik/studio A3") 40 hours, Device B (Samsung Galaxy S7), Device C (iPhone SE) 3h each. Additionally, 6 different microphones are simulated, 3 hours each.
We choose the training/testing setup to test (a) Generalization to unseen cities (b) Generalization to unseen devices. 
We ensure that each (city, class) pair appears only in either the train or the test set. We train using only recordings from Device A, and test both on recordings from Device A (referred to as ``Seen'') and on recordings from the other devices (referred to as ``Unseen''). In both cases, we only test on unseen cities for an acoustic scene.




\section{Creating Over-Parameterized Models}

\label{sec:network}
In our experiments, we use a baseline architecture, namely, a variant of the ResNet~\cite{heDeepResidualLearning2016} proposed in~\cite{Koutini2019Receptive}, which is the state-of-the-art in low-complexity ASC under device mismatch, on the DCASE benchmark~\yrcite{Heittola2020}.
The configuration of the network is detailed in Table~\ref{tab:arch}, where the width, i.e., number of channels, of each layer is controlled by $W$, and $D,K$ control the depth of the network. 
Using this architecture, a baseline network with $W=32, D=1, K=1$ can already fit the training data with $99.99\%$ accuracy. Furthermore, this baseline can fit the training data with random labels to $85.51\%$ accuracy. 
The same network with $W=64$ fits the training data with random labels to $100\%$ accuracy, indicating model over-parameterization~\cite{rethinkinggeneralizationZhangBHRV17}.  
In addition, we compare the network in each setup to the \emph{Frequency-Damped} version of our baseline architecture~\cite{koutini21journal}. 
Frequency-damping reduces the Effective Receptive Field (ERF)~\cite{luoUnderstandingEffectiveReceptive2016} of the network over the frequency dimension, by down-scaling the convolutional weights away from the center of the RF along the frequency dimension, as explained in Appendix~\ref{appdx:damped}.
Frequency-Damped networks have an inductive bias towards limiting the RF, which is beneficial for audio tasks and have shown success in the acoustic domain~\cite{koutini21journal}.
We refer to this variant as \emph{Damped}, and use it to further test the importance of task-specific bias/regularization under different over-parameterization settings.

\begin{table}[t]
\caption{Baseline architecture and scaling factors}
\label{tab:arch}
\vskip 0.15in
\begin{center}
\begin{small}
\begin{sc}
\begin{tabular}{lccr}
\toprule
\textbf{Repeat}&\textbf{Channels}&\textbf{Block}&\textbf{Config} \\
\midrule
& $W$ & Input & $ 5 \times 5$ stride=$2$ \\
\midrule
1&$W$ & R&$3 \times 3$, $ 1 \times 1$, P\\
1&$W$  & R  & $ 3 \times 3$,  $  3 \times 3$, P  \\
1&$W$  & R  & $ 3 \times 3$,  $ 3 \times 3$  \\
1&$W$  & R    & $ 3 \times 3$,  $ 3 \times 3$, P   \\
\midrule
$D$ &$2 \times W$  &R& $3 \times 3$, $ 3 \times 3 $  \\
\midrule
$K$&$4 \times W$  &R & $ 1 \times 1$, $ 1 \times 1$  \\
\midrule
\multicolumn{4}{c}{Classifier $4 \times W \rightarrow 10 $ classes }   \\ 
\multicolumn{4}{c}{Global mean pooling }   \\ 
\bottomrule
\multicolumn{4}{l}{ P: $ 2 \times 2$ max pooling.}\\
\multicolumn{4}{l}{ R: Residual, the input is added to the output}\\
\end{tabular}
\end{sc}
\end{small}
\end{center}
\vskip -0.1in
\end{table}

We scale the baseline networks by increasing the width of the CNN, choosing $W$ (Table~\ref{tab:arch}) in $\{ 32, 64, 128, 192 ,256, 384, 512\}$. Results are discussed in Section~\ref{subsec:width}.
For comparison, we scale the networks by increasing the depth, adding residual blocks with $ 3 \times 3 $ convolutions (increasing $D$ in Table~\ref{tab:arch}). This will result in networks with more parameters, and bigger RFs over the input.
However, since such increasing of depth has been shown to cause overfitting on this task~\cite{koutini21journal}, we additionally investigate increasing the depth without increasing the RF of the networks, by adding residual blocks with $ 1 \times 1 $ convolutions (increasing $K$ in Table~\ref{tab:arch}). 
The depth scaling results are presented in Section~\ref{subsec:depth}.

We also study whether the observed performance improvement is due to increased over-parameterization, or due to widening of the network.
To that end, we use the same setup as~\citet{golubeva2020wider}, and randomly mask out weights of the network at initialization, prioritizing those layers with more parameters, and increase sparsity of the weights. We refer to this approach as \emph{static} random pruning.
As a result, we maintain the same number of parameters as the network before widening. We use the number of weights of the network with $W=64$ as a base (since it can fit the random labels to $100\%$ accuracy), and sparsify the larger networks to have the same number of weights. We further compare this method to \emph{iteratively} and \emph{selectively} removing the lowest-magnitude weights while training. \citet{frankle2021pruingatinit} show that the latter is expected to have better performance than random static pruning~\cite{golubeva2020wider}, as well as other more complex methods for pruning at initialization, such as SynFlow~\cite{TanakaKYG20synflow} and SNIP~\cite{LeeAT19SNIP}. Although iterative magnitude pruning allows the network to have more parameters during (parts of the) training, at evaluation all the networks have the same number of parameters.  We present the results of sparse wide networks in Section~\ref{subsec:sparse_width}.

\begin{figure}[t]
\begin{center}
\centerline{\includegraphics[width=\columnwidth]{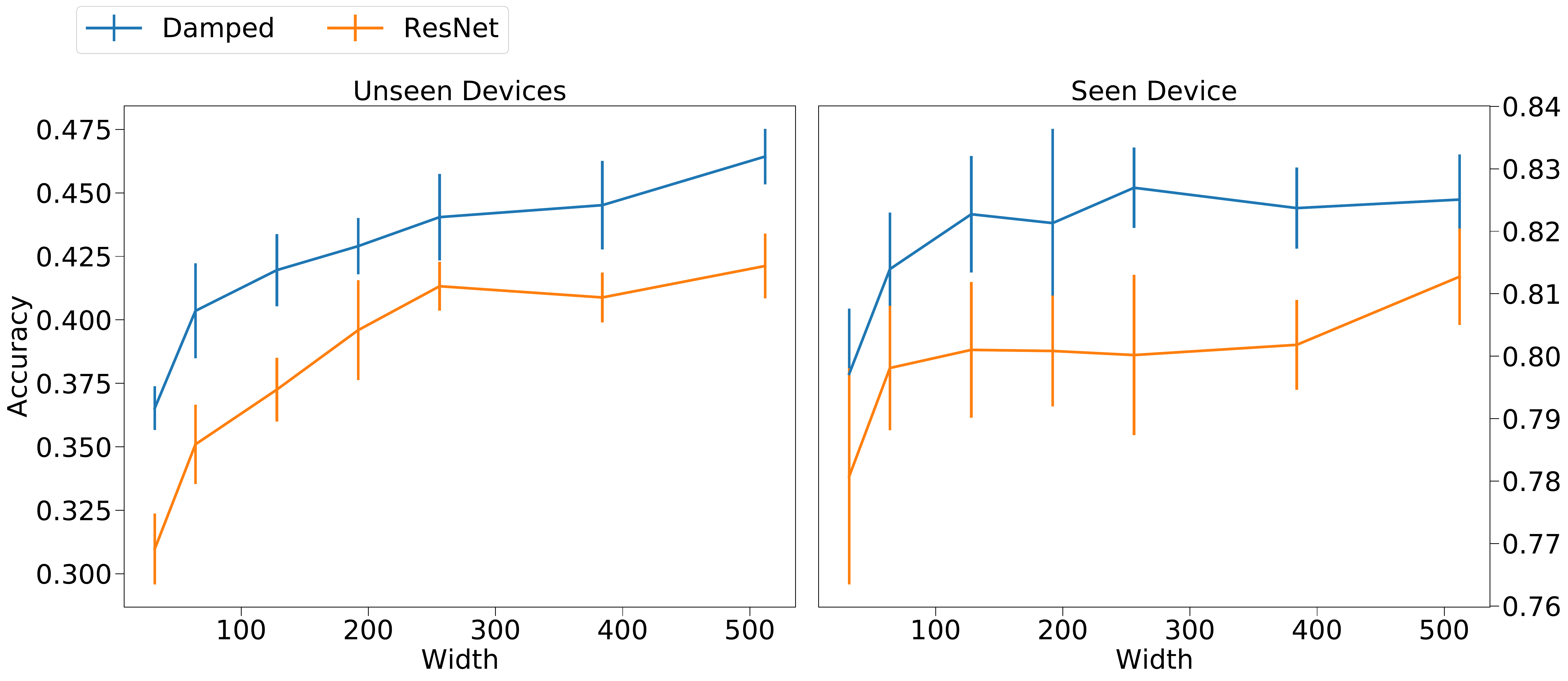}}
\caption{The effect of the width on accuracy.}
\label{fig:seen:width}
\end{center}
\vskip -0.2in
\end{figure}

\section{Results and Discussion}

Figure~\ref{fig:summery:damped} shows a summary of the accuracy as a function of the number of parameters for the different scaling methods for the Damped version.
As can be seen, models scaled up in their parameters by increasing width have significantly better generalization on unseen devices compared to models that are scaled up using depth (with, or without increasing the RF).
Furthermore, the generalization capabilities of models scaled up with width and depth are on par with the best-case scenarios.

\subsection{Increasing the Width}
\label{subsec:width}
Figure~\ref{fig:seen:width} (right) shows that the performance on the seen device improves initially by increasing the network's width. 
Increasing $W$ to values larger than $128$ seems to have minimal effect on the performance. 
Looking at the results on unseen devices (left), we can observe that widening both architectures significantly improves their accuracy.
Compared to the baseline ($W=32$, which can fit the training data to 99.99\% accuracy), setting $W=512$ results in 10\% increase in accuracy on unseen devices for both architectures.
Additionally, the Damped model, which encodes the task-specific bias, outperforms the base network in all settings.

\subsection{Increasing the Depth}
\label{subsec:depth}

\begin{figure}[t]
\begin{center}
\centerline{\includegraphics[width=\columnwidth]{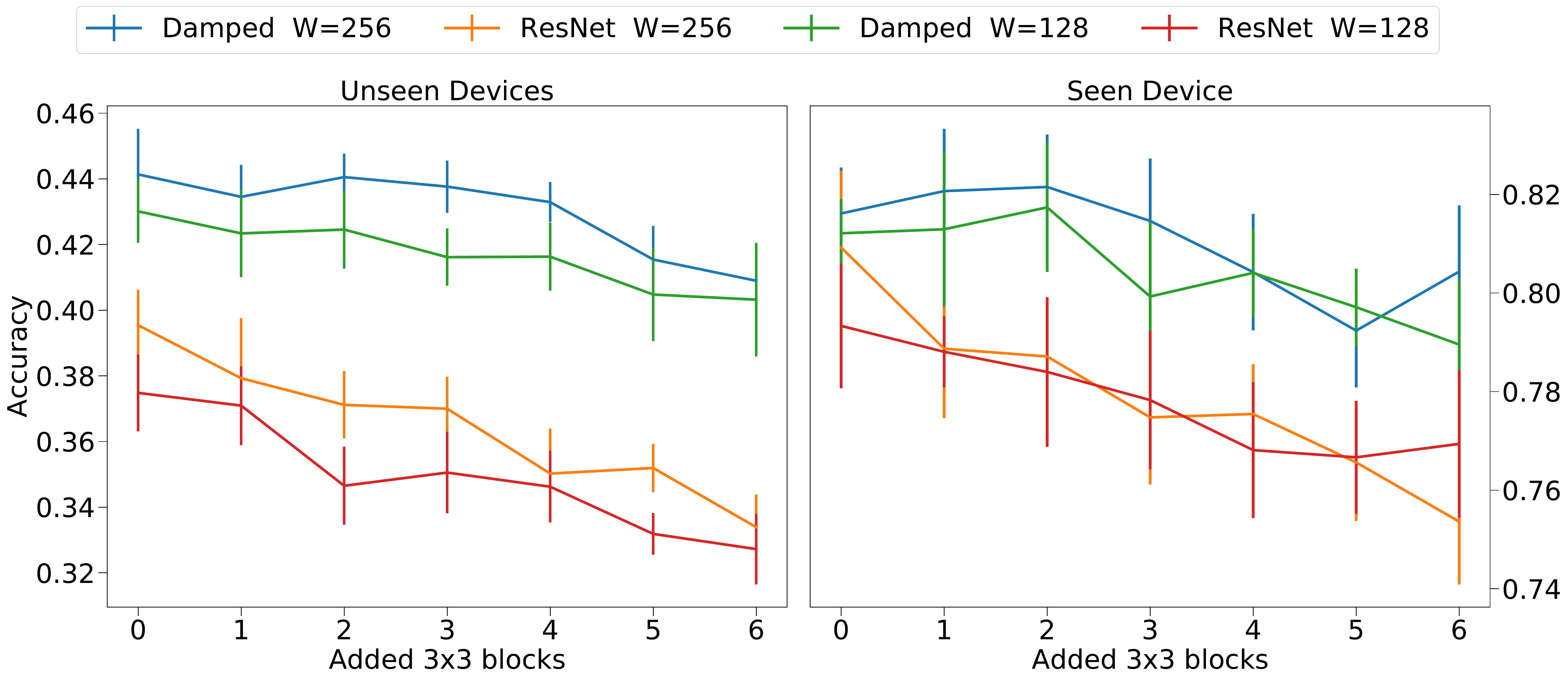}}
\caption{The effect of the depth on accuracy.}
\label{fig:seen:depth}
\end{center}
\vskip -0.2in
\end{figure}

\begin{figure}[t]
\begin{center}
\centerline{\includegraphics[width=\columnwidth]{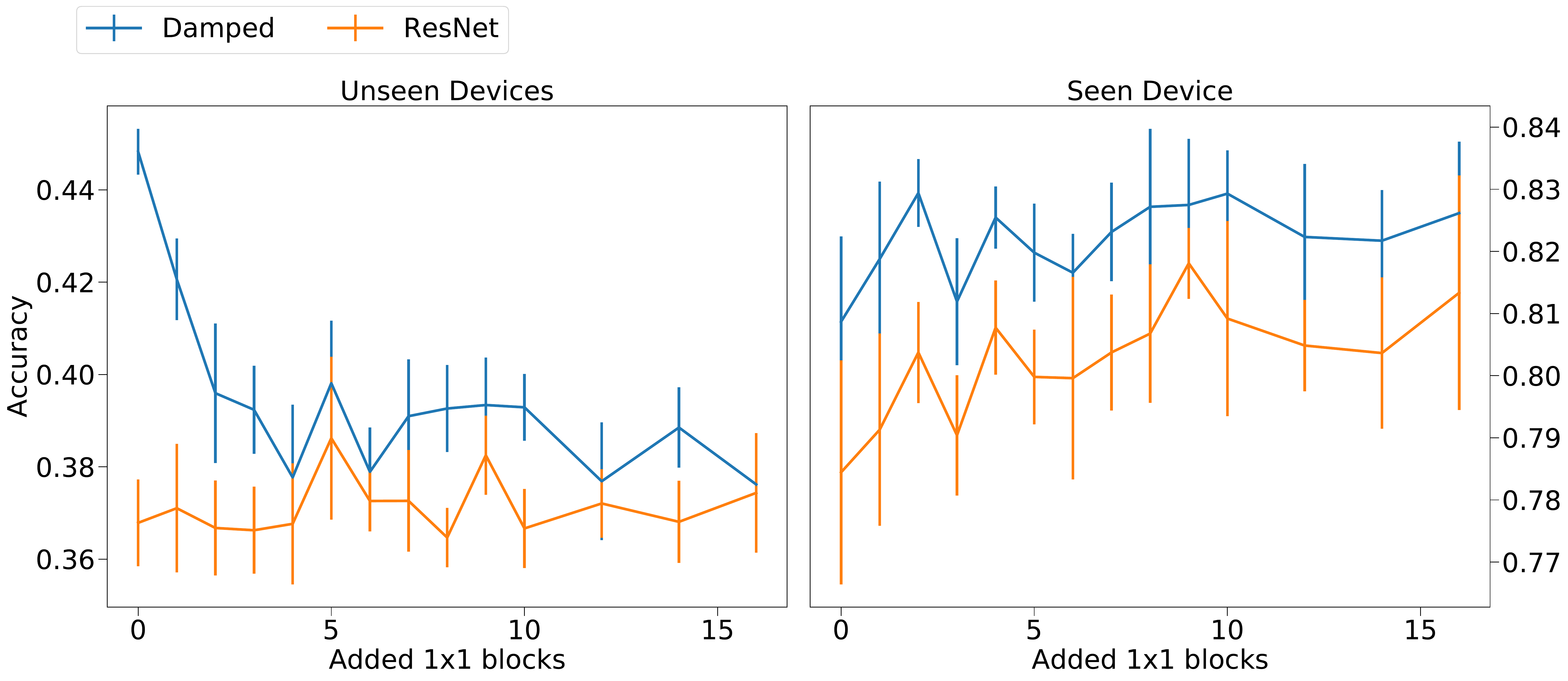}}
\caption{The effect of the depth (without increasing RF) on Accuracy.}
\label{fig:seen:1b1depth}
\end{center}
\vskip -0.2in
\end{figure}

\begin{figure}
\begin{center}
\centerline{\includegraphics[width=0.7\columnwidth]{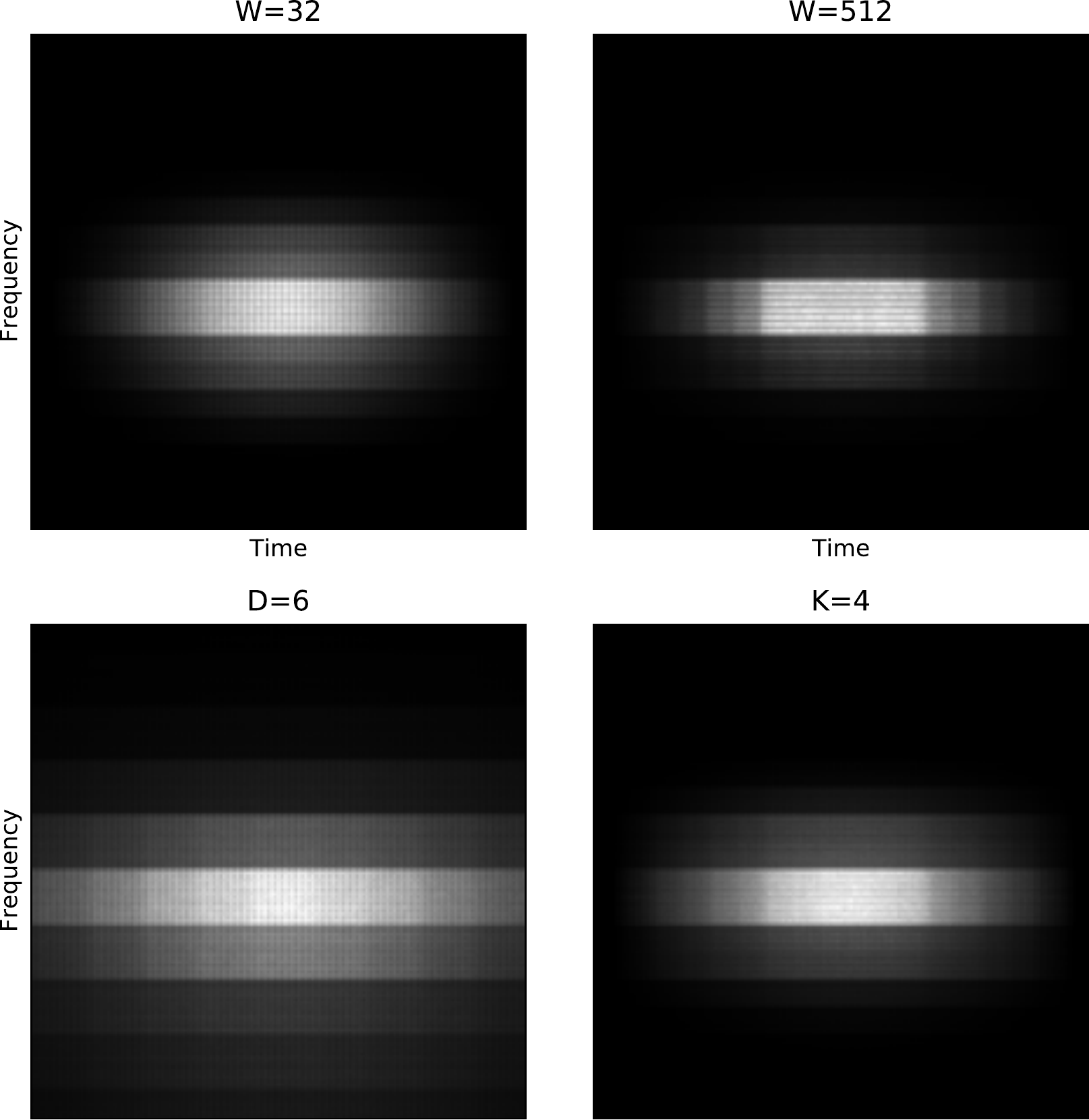}}
\caption{The ERF of different scaled variants of the Damped ResNet. The figures are cropped to the Max RF of the network with $D=1$ for visibility.}
\label{fig:erf_all_compare}
\end{center}
\vskip -0.3in
\end{figure}

Figure~\ref{fig:seen:depth} shows the performance of the networks on seen and unseen devices when depth is increased. 
In both cases, the experiments show that increasing the depth negatively affects the performance. This effect is larger on the base network compared to the damped variant. 
\citet{koutini21journal} explain this by the increase in the RF of the network, which is not suitable for audio classification tasks. 
Figure~\ref{fig:seen:1b1depth} shows that scaling the depth with $1\times 1 $ convolutions -- which does not change the maximum RF of the network -- can give small improvements on the seen device, this effect does not translate to the unseen devices. 
Moreover, adding these layers worsens the performance of the damped model. 
Looking at Figures~\ref{fig:erf_all_compare} and ~\ref{fig:depth_erf_comapre}, we can explain this by the observation that adding additional $1\times 1 $ convolutions increases the ERF, countering the effect of damping (also see Appendix~\ref{appdx:erf_depth_damped}).

\subsection{Increasing the Width with Sparsification}
\label{subsec:sparse_width}

\begin{figure}
\begin{center}
\centerline{\includegraphics[width=\columnwidth]{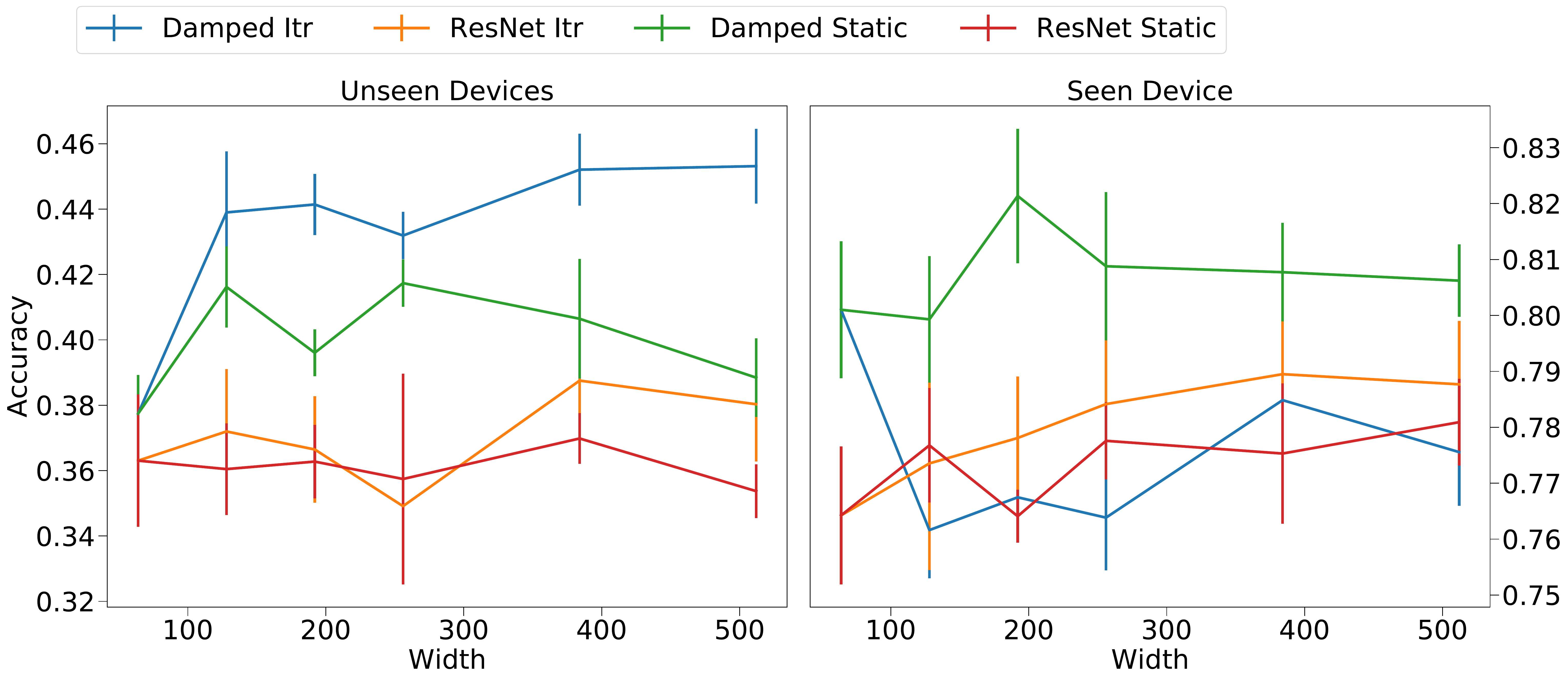}}
\caption{The effect of the \emph{sparse} width on the accuracy. All the networks have the same number of parameters. Static: random masking at initialization. Itr: iterative magnitude pruning.}
\label{fig:wide_sparse}
\end{center}
\vskip -0.2in
\end{figure}

Figure~\ref{fig:wide_sparse} shows the effect of widening the networks while keeping the number of parameters equivalent to the model with $W=64$. Using static random pruning at initialization, we see initial performance improvements with increasing width. However, these gains diminish on wider networks, in accordance with the findings of~\citet{golubeva2020wider}. 
Surprisingly, the performance of the damped version drops on the seen device in iterative pruning. The ERF plots shown in Figure~\ref{fig:width_erf_comapre:itr} provide a possible explanation: the damped weights are more likely to be pruned first, since they have a lower magnitude, resulting in a smaller ERF.
On the contrary, the performance suffers a minimal drop on the unseen devices, even for the model with $W=512$ that has $98.4\%$ sparsity.

\section{Conclusion}
We study different methods to scale up CNNs in ASC, and test their ability to generalize to unseen devices. 
Our experiments show that scaling up the width of CNNs plays the biggest role in their generalization to unseen devices, even if no gains are observed on the already seen devices.
We further show that widening the network without increasing parameters 
can still improve generalization on unseen devices.
Additionally, the results suggest that depth scaling can lead to worse performance on unseen devices.
Finally, we show a possible interpretation of the results in terms of the ERF of the network.




\section*{Acknowledgements}

This work has been supported by the COMET-K2 Center of the Linz Center of Mechatronics (LCM) funded by the Austrian Federal Government and the Federal State of Upper Austria.
The LIT AI Lab is financed by the Federal State of
Upper Austria.

\section*{Software and Data}
The source code is available on \url{https://github.com/kkoutini/cpjku_dcase21}.


\bibliography{refs}
\bibliographystyle{icml2021}

\appendix
\section{Experiment Setup}
\label{sec:exp_setup}

\subsection{Preprocessing}
The input clips are $10$ seconds, and have a sampling rate of $22050$ Hz. We apply Short Time Fourier Transform (STFT) on the input audio signals, with a window size of $2048$ and stride of $512$. We use perceptual weighting to scale the resulting frequencies, and apply mel filter bank with $256$ filters. This results in an input for the CNNs of the shape $256 \times 431$ .

\subsection{Augmentation}
We roll the raw audio wave over time, and use \textit{Mix-up}~\cite{zhangMixupEmpiricalRisk2017} augmentation with $\alpha=0.3$ , as it has been shown to improve performance on this task~\cite{koutini21journal}.

\subsection{Training}
We train for 250 epochs using Adam~\cite{kingmaAdamMethodStochastic2014} with a max learning rate of $10^{-4}$, starting with $20$ epochs of warm-up by exponentially ramping up the learning rate, followed by a linear learning rate decay to $10^{-6}$. We repeat all the experiments $3$ times and report the mean the standard deviation of the last $10$ epochs of the runs.

\subsection{Sparsification}
In Sparsification experiments, we use 350 epochs of training. For static random pruning, we use the method proposed by~\citet{golubeva2020wider} 
to remove weights from the layers with the largest number of parameters first. In iterative pruning, the number of weights removed in each epoch decays exponentially until we reach the desired number of parameters. Since we are doing global pruning, we prevent layer collapse, by excluding the layers which have more than $99\%$ sparsity from further pruning. Without this measure, we observe significant performance drops starting from $W=192$.

\section{Frequency-Damped Convolutions}
\label{appdx:damped}

\begin{figure}[h]
\begin{center}
\centerline{\includegraphics[width=\columnwidth]{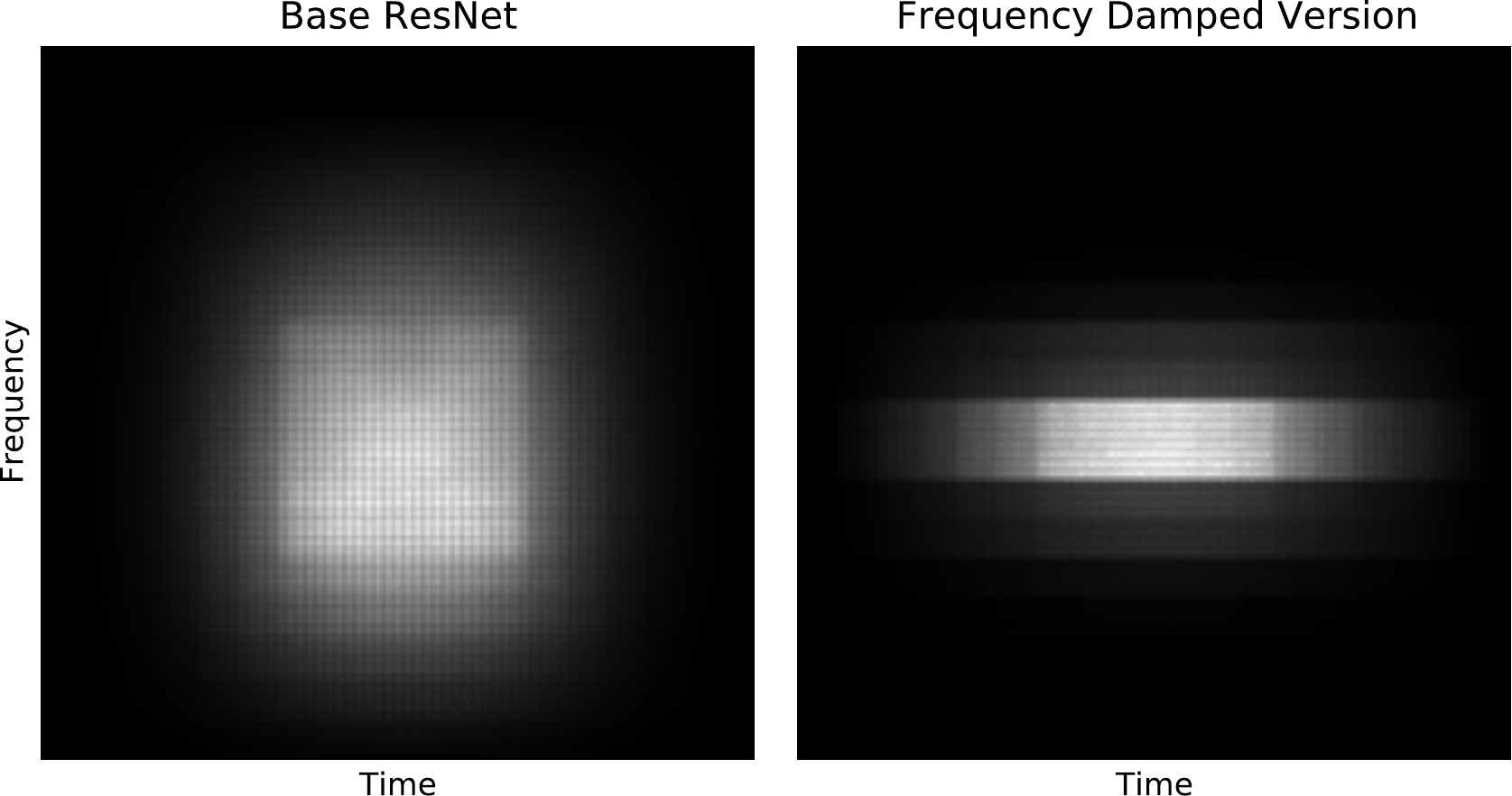}}
\caption{The \emph{effective receptive field} of a ResNet with $W=128, D=1, K=1$ and it's damped version.}
\label{fig:erf_damped_compare}
\end{center}
\vskip -0.2in
\end{figure}

Damped convolutions~\cite{koutini21journal} encodes an inductive bias for audio spectrograms, by restricting the Effective Receptive Field (ERF)~\cite{luoUnderstandingEffectiveReceptive2016} of the CNNs over the input spectrograms. 
In this paper, we use \emph{frequency-damped convolution}
~\yrcite{koutini21journal} by replacing every convolutional operator in the  $O_n  = W_n \star Z_{n-1} + B_n$ with $O_n = (W_n \odot C_n) \star Z_{n-1} + B_n$, where $\star$ is the convolution operator, $\odot$ is the element-wise multiplication, $Z_n$ is the output of the nth layer. $C_n$ is a constant matrix, which has the same shape as the filter weights $W_n$, $C_n$ has a value of 1 in the center and decays away from the center on the frequency dimension. This biases the classification layer to fit the center of its receptive field, and makes fitting the outer region of its maximum receptive field harder. Figure~\ref{fig:erf_damped_compare} shows the difference in the effective receptive field between a ResNet and its damped version.

\section{The Effect of Scaling on the Effective Receptive Field}
We calculate the \emph{effective receptive field} as proposed by \citet{luoUnderstandingEffectiveReceptive2016}, by calculating the input  gradients to a change in the activation of a single spatial pixel of the penultimate layer of the network (We choose the middle pixel in all the illustrated plots). We take the mean of the absolute gradients over all the test data. The plots show the mean of the absolute gradients over the whole unseen test set. The plots are cropped to the maximum receptive field of the CNN (which can be calculated from the filters' sizes and the strides of the CNN layers) for better visibility. 

\subsection{Depth}
\label{appdx:erf_depth_damped}
We show the \emph{ERF} of a Frequency-Damped ResNet with $W=128, D=1, K=0$ and depth-scaled version without increasing the maximum receptive field, by stacking 4 more blocks with filters sizes of $1 \times 1$. The figure shows that adding these blocks -- although doesn't change the maximum receptive field -- increases the effective receptive field, effectively countering the effect of damping. 

\begin{figure}[h]
\begin{center}
\centerline{\includegraphics[width=\columnwidth]{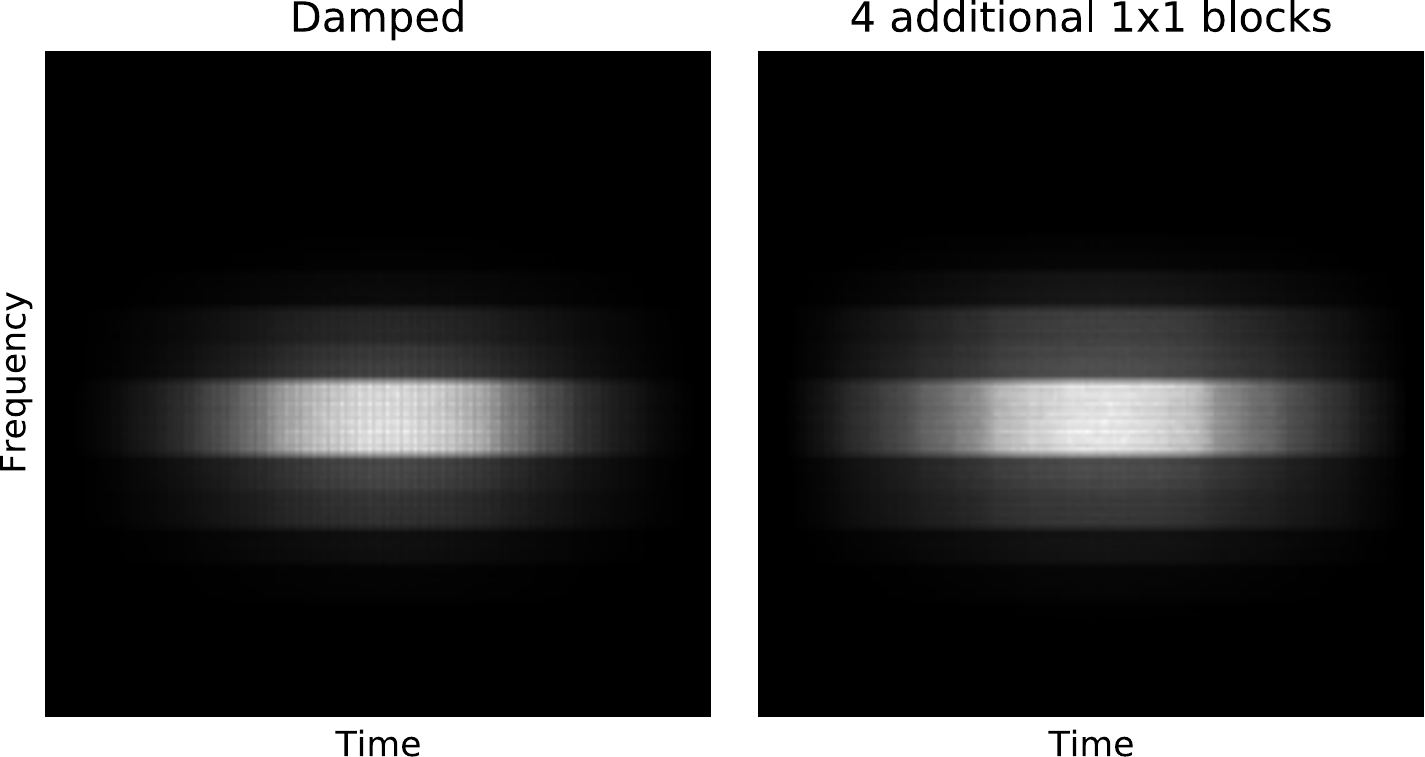}}
\caption{The \emph{effective receptive field} of a Frequency-Damped ResNet with $W=128, D=1, K=0$ and $K=4$.}
\label{fig:depth_erf_comapre}
\end{center}
\end{figure}

\begin{figure}[ht!]
\begin{center}
\centerline{\includegraphics[width=\columnwidth]{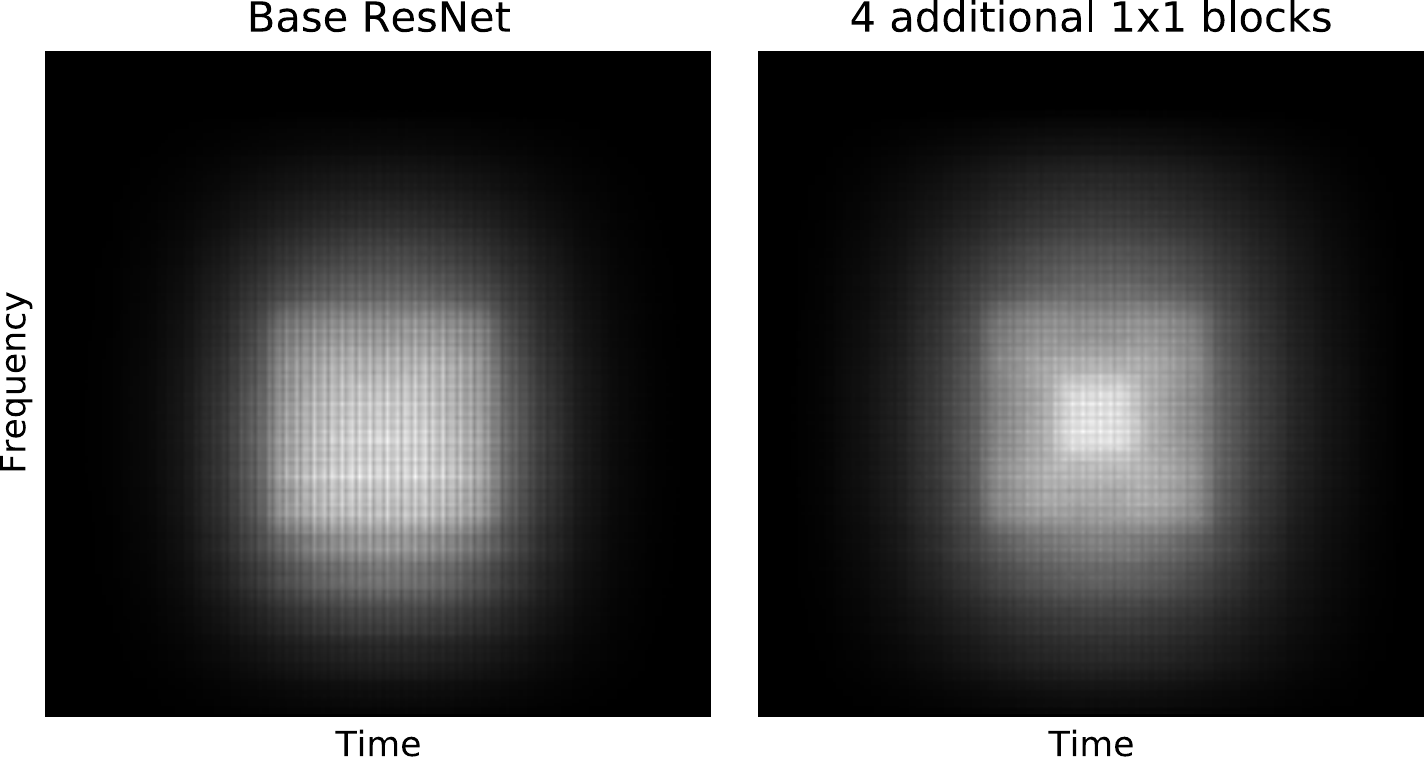}}
\caption{The \emph{effective receptive field} of the Base ResNet with $W=128, D=1, K=0$ and $K=4$.}
\label{fig:depth_erf_comapre:resnet}
\end{center}
\end{figure}


\subsection{Width}
\label{appdx:erf_width_damped}
\begin{figure}[h!]
\begin{center}
\centerline{\includegraphics[width=\columnwidth]{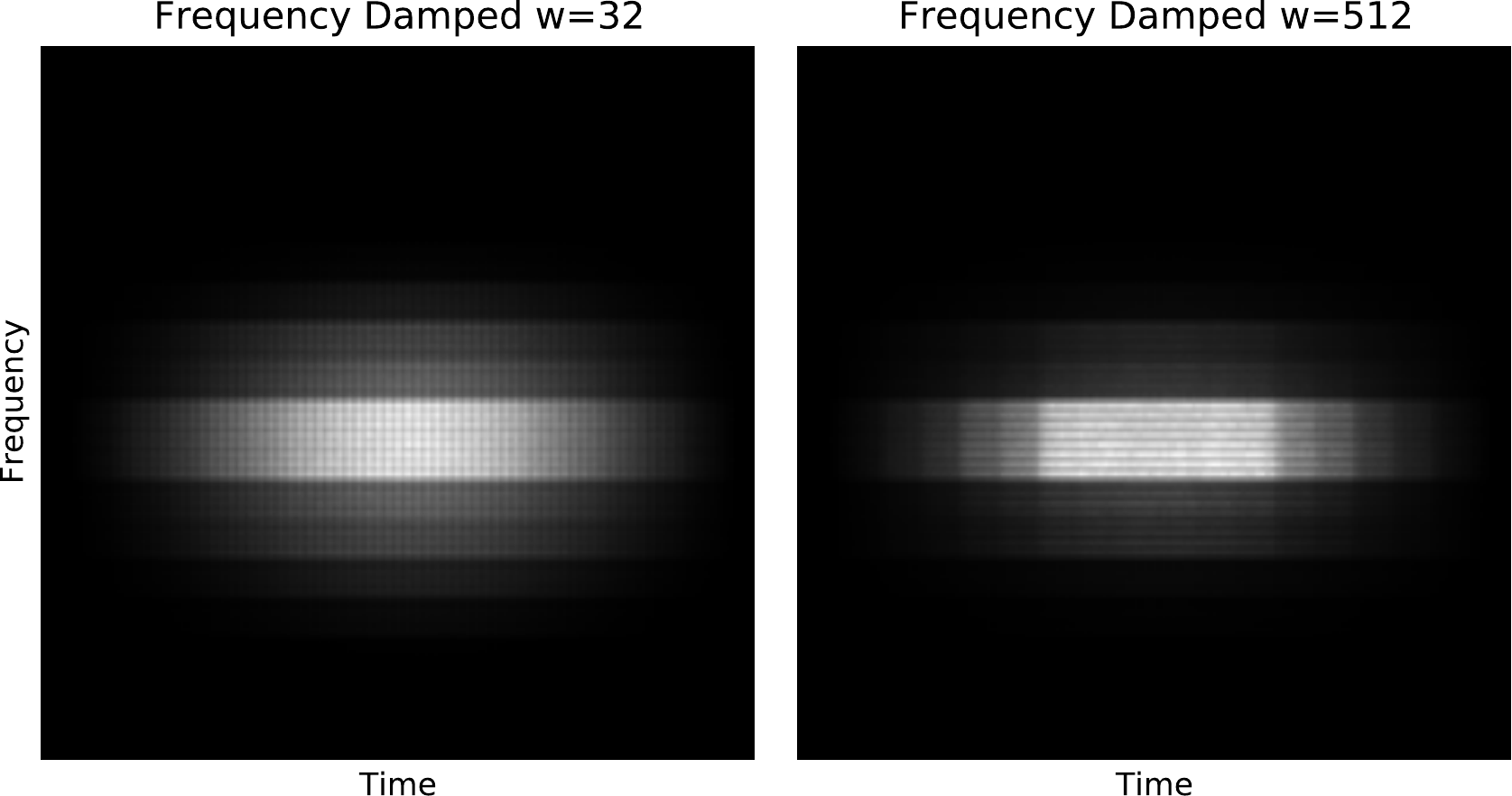}}
\caption{The \emph{effective receptive field} of a Frequency-Damped ResNet with $W=32, D=1, K=0$ and $W=512, D=1, K=0$.}
\label{fig:width_erf_comapre}
\end{center}
\end{figure}

\begin{figure}[h]
\begin{center}
\centerline{\includegraphics[width=\columnwidth]{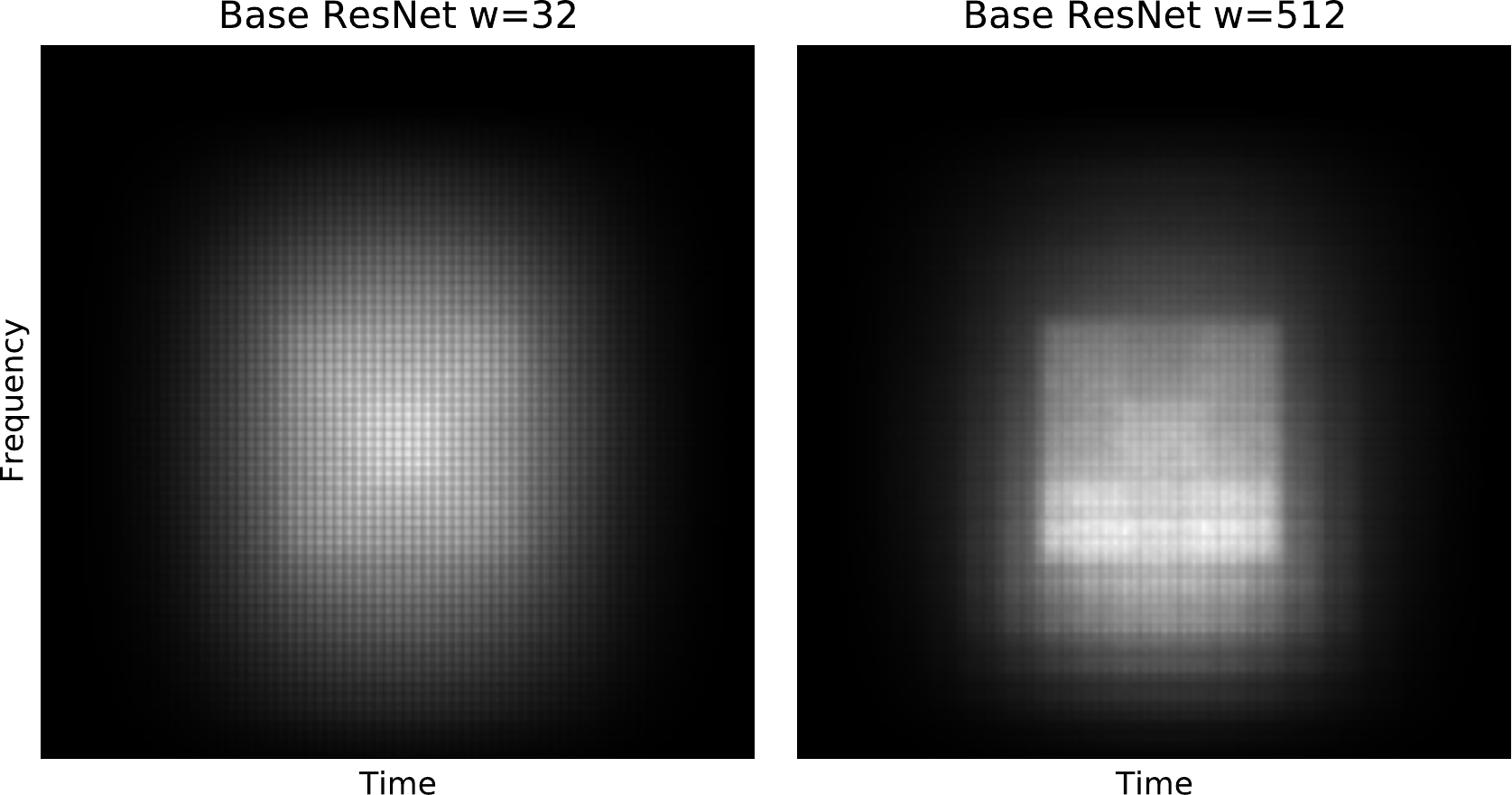}}
\caption{The \emph{effective receptive field} of the base ResNet with $W=32, D=1, K=0$ and $W=512, D=1, K=0$.}
\label{fig:width_erf_comapre_res}
\end{center}
\end{figure}

Figures~\ref{fig:width_erf_comapre} and Figures~\ref{fig:width_erf_comapre_res} show that increasing the width of both variants focuses the ERF in the center of the Max RF. 

\subsection{Width with Pruning}
\label{appdx:erf_width_damped:prune}
\begin{figure}[h]
\begin{center}
\centerline{\includegraphics[width=\columnwidth]{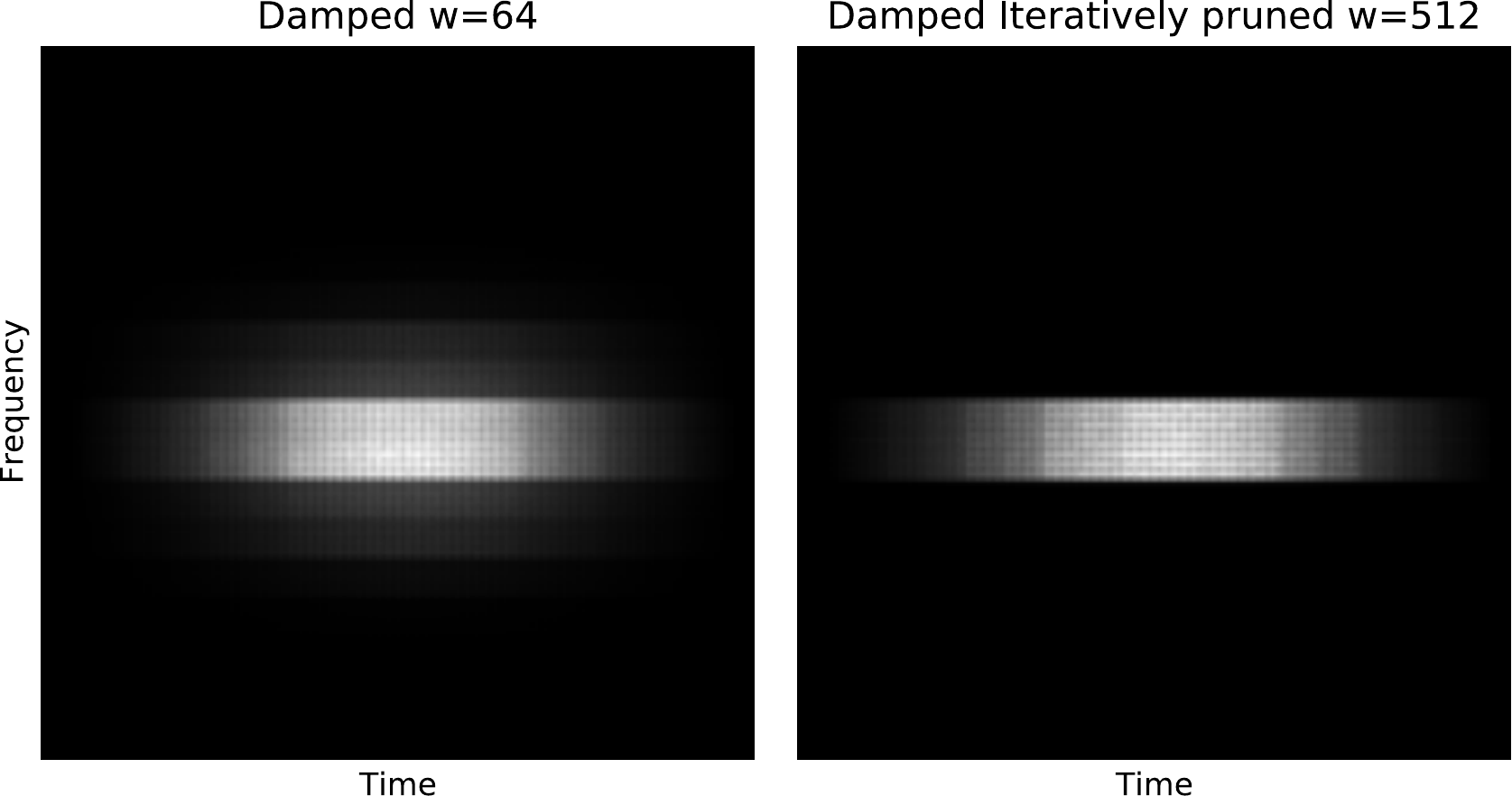}}
\caption{The \emph{effective receptive field} of a Frequency-Damped ResNet with $W=64, D=1, K=0$ and $W=512, D=1, K=0$ with iterative magnitude pruning. Both networks have the same number of parameters.}
\label{fig:width_erf_comapre:itr}
\end{center}
\end{figure}

\begin{figure}[h]
\begin{center}
\centerline{\includegraphics[width=\columnwidth]{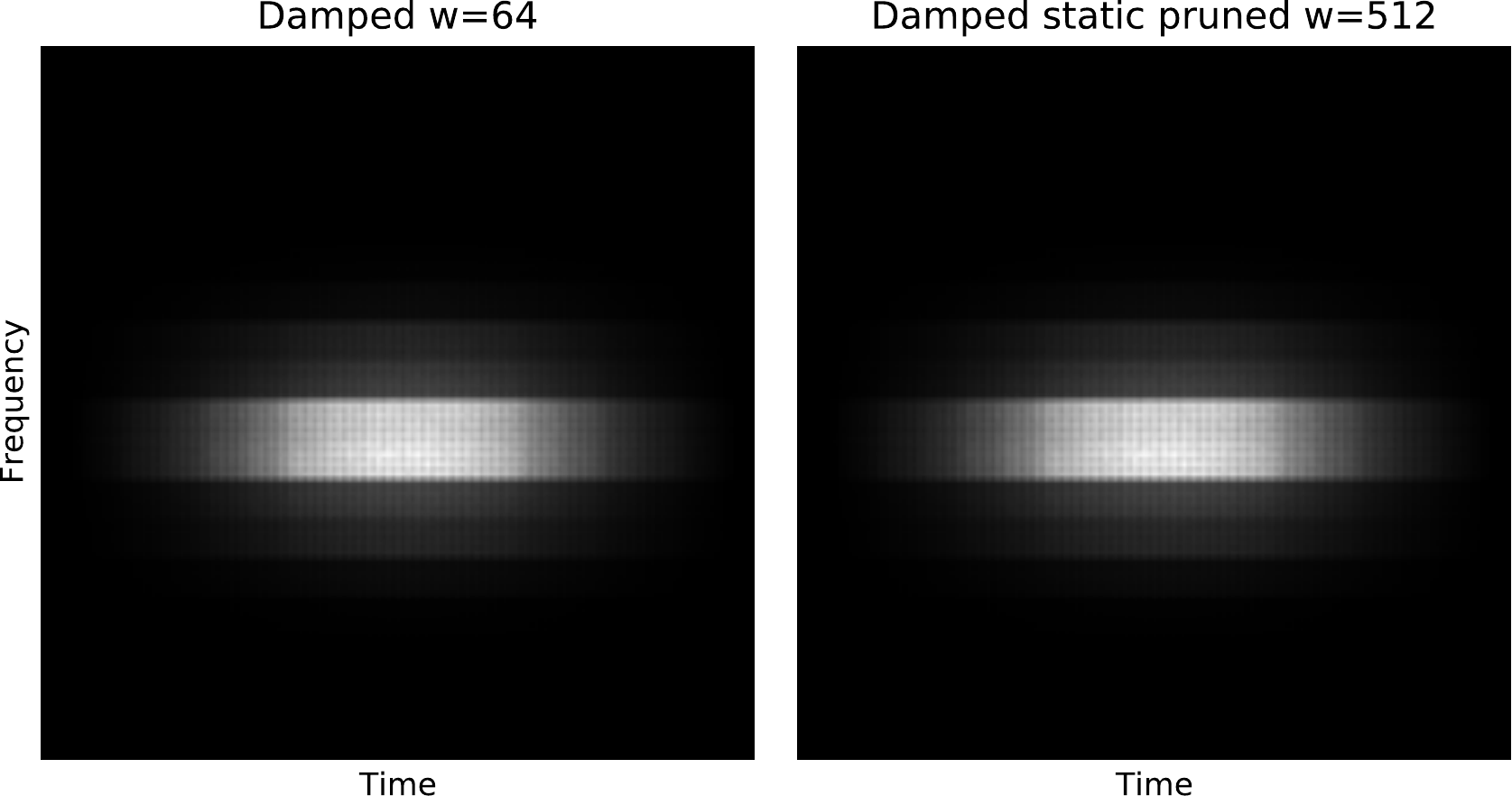}}
\caption{The \emph{effective receptive field} of a Frequency-Damped ResNet with $W=64, D=1, K=0$ and $W=512, D=1, K=0$ with random pruning at initialization. Both networks have the same number of parameters.}
\label{fig:width_erf_comapre:randprune}
\end{center}
\end{figure}

\begin{figure}[h]
\begin{center}
\centerline{\includegraphics[width=\columnwidth]{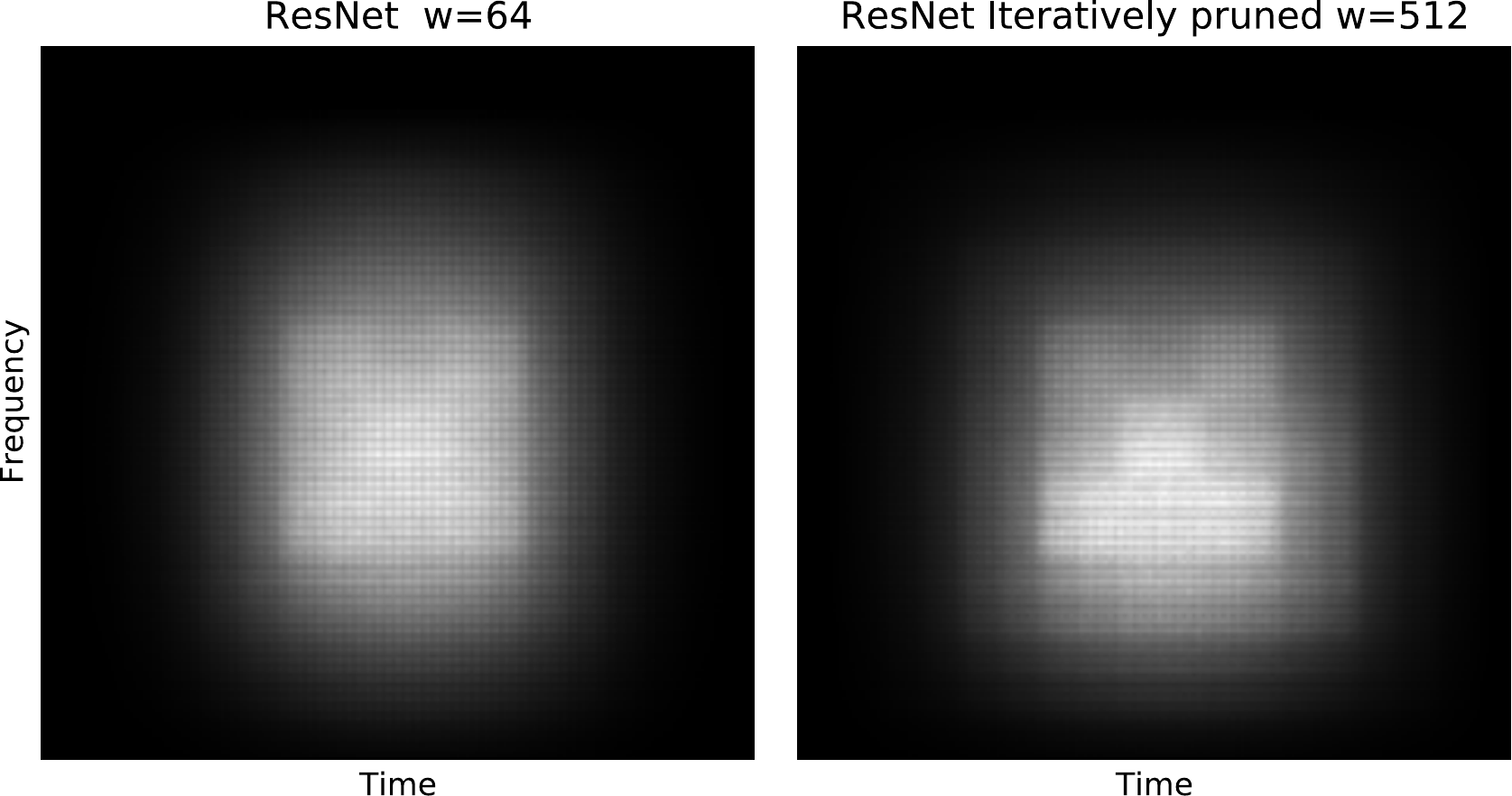}}
\caption{The \emph{effective receptive field} of the base ResNet with $W=64, D=1, K=0$ and $W=512, D=1, K=0$ with iterative magnitude pruning. Both networks have the same number of parameters. }
\label{fig:width_erf_comapre:res:itr}
\end{center}
\end{figure}

Figure~\ref{fig:width_erf_comapre:itr} shows that magnitude pruning reduces the ERF size over frequency for the Damped ResNet, since favors the damped weights (since they are down-scaled). This does not happen in random pruning at initialization, Figure~\ref{fig:width_erf_comapre:randprune}. On the base ResNet, comparing Figures~\ref{fig:width_erf_comapre:res:itr} and  Figure~\ref{fig:width_erf_comapre_res} shows widened networks have a similar ERF with and without iterative pruning. 

\begin{figure}[h]
\begin{center}
\centerline{\includegraphics[width=\columnwidth]{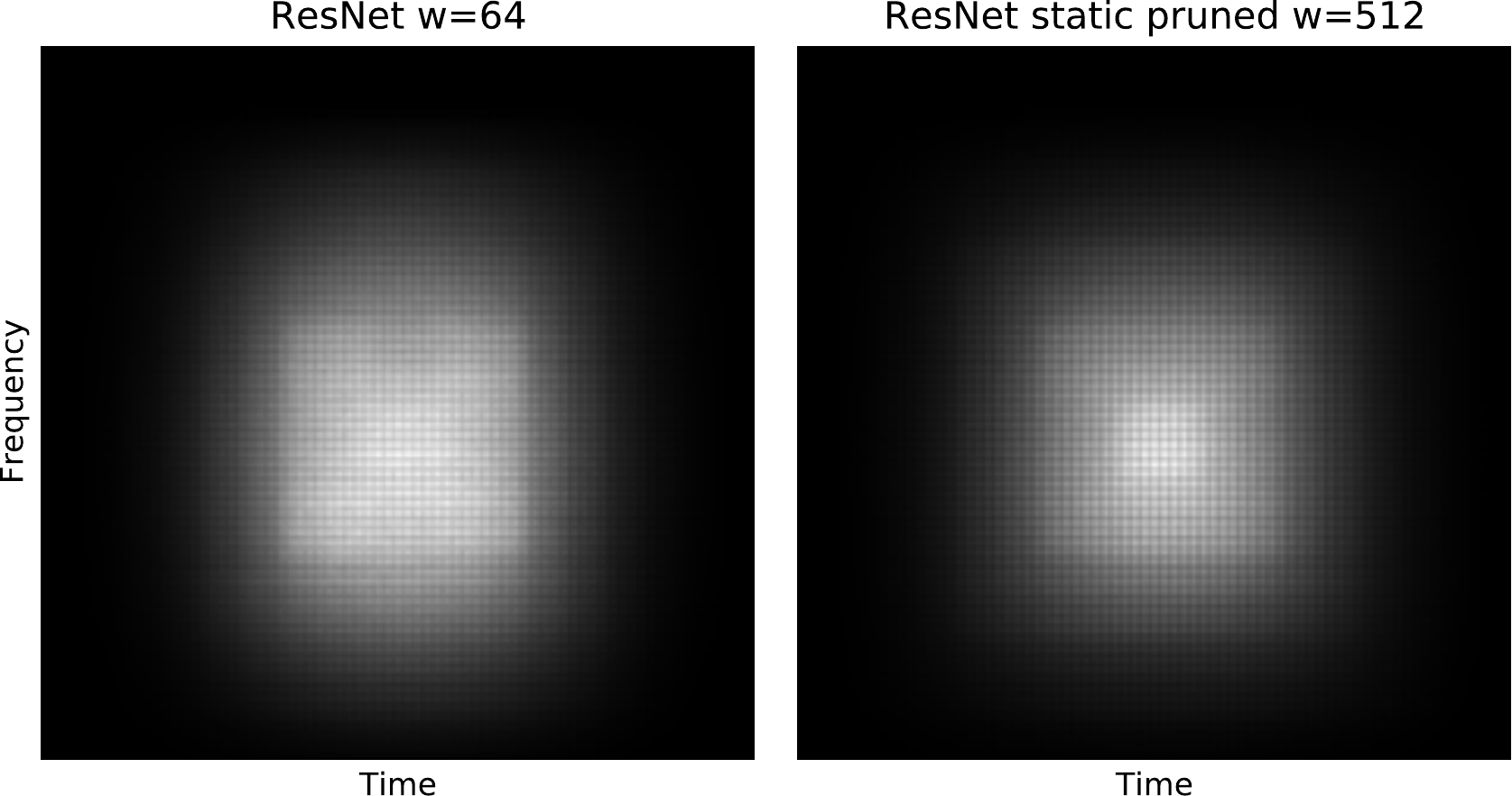}}
\caption{The \emph{effective receptive field} of the base ResNet with $W=64, D=1, K=0$ and $W=512, D=1, K=0$ with with random pruning at initialization. Both networks have the same number of parameters. }
\label{fig:width_erf_comapre:res:randprune}
\end{center}
\end{figure}

\begin{figure}[h]
\begin{center}
\centerline{\includegraphics[width=0.95\columnwidth]{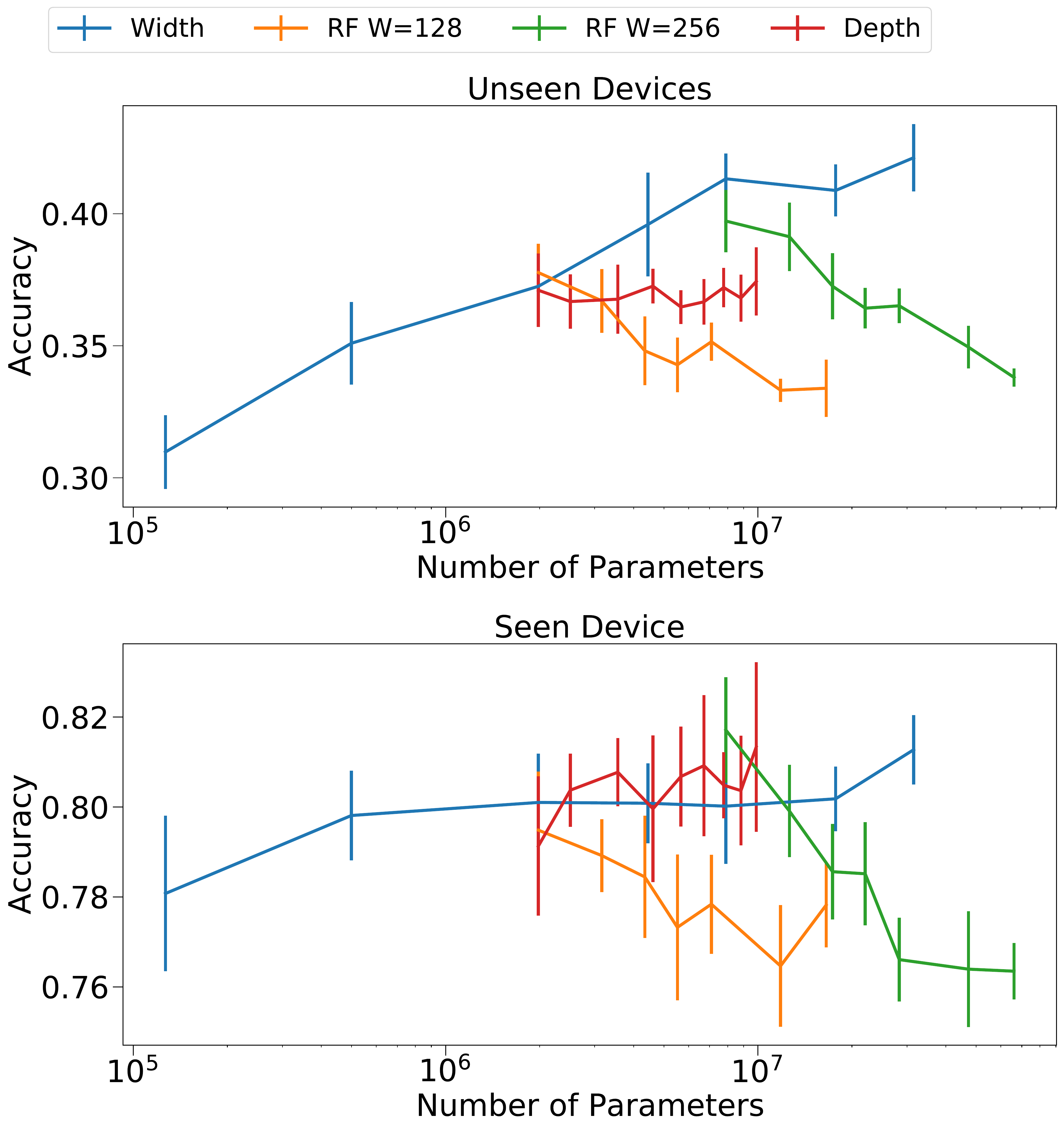}}
\caption{Test accuracy of scaled up variants of the Base ResNet as the number of parameters increase, as explained in Section~\ref{sec:network}.  }
\label{fig:summery_resnet}
\end{center}
\end{figure}


\end{document}